\documentclass{aa}
\usepackage{graphics}
\usepackage{times}

\newcommand{\grs}   {GRS 1915+105}

\newcommand{\rmag}   {${\rlap.}^{m}$}

\newcommand{\ltsima} {$\; \buildrel < \over \sim \;$}
\newcommand{\simlt}  {\lower.5ex\hbox{\ltsima}}            
\newcommand{\gtsima} {$\; \buildrel > \over \sim \;$}
\newcommand{\simgt}  {\lower.5ex\hbox{\gtsima}}            

\begin{document}

\thesaurus{08(08.09.2 \object{\grs}; 13.09.6; 13.18.5; 13.25.5)}

\title{VLT observations of \grs}

\author{Josep Mart\'{\i}\inst{1}
\and    I. F\'elix Mirabel\inst{2,3} 
\and    Sylvain Chaty\inst{4}
\and    Luis F. Rodr\'{\i}guez\inst{5}
}

\institute{
Departamento de F\'{\i}sica, Escuela Polit\'ecnica Superior,
Universidad de Ja\'en, Calle Virgen de la Cabeza, 2, E-23071 Ja\'en, Spain
\and
CEA/DSM/DAPNIA/Service d'Astrophysique, Centre d'\'Etudes de Saclay,
F-91191 Gif-Sur-Yvette, France
\and
Instituto de Astronom\'{\i}a y F\'{\i}sica del Espacio, C.C. 67, Suc. 28, 1428 Buenos Aires, Argentina
\and
Physics Department, The Open University, Walton Hall, Milton Keynes, MK7 6AA, United Kingdom 
\and
Instituto de Astronom\'{\i}a, UNAM, Apdo. Postal 70-264, 04510 M\'exico D.F., M\'exico
}

\offprints{J. Mart\'{\i}, jmarti@ujaen.es}

\date{Received / Accepted}

\maketitle

\begin{abstract}

We present near infrared spectroscopy of the superluminal
microquasar \grs\ obtained with the first unit of the VLT \footnote{Based on observations collected
at the European Southern Observatory, Chile (ESO No 63.H-0261).}
and the ISAAC spectro-imager.  The emission features detected in the
VLT data have been identified as He I, Br $\gamma$, He II and Na I.
The detection of Na I is reported here for the first time, while 
our confirmation of weak He II emission provides support to previous marginal detections of this feature. 
By comparing the observed spectra with those of massive stars,
we find that our results are very consistent with \grs\ being
a high mass X-ray system with
an early type primary, as previously proposed by Chaty et al. (1996) and  Mirabel et al. (1997). 
The VLT spectra also provide evidence of P Cygni profiles, that turn into
blue emission wings when the system is in outburst. This observed line profile evolution  
implies that \grs\ must be surrounded by an expanding envelope,
that is partially blown out during the X-ray outbursts. The presence of such
circumstellar gaseous material around \grs\ is more naturally understood in the context
of a massive luminous star than if the system was a low-mass X-ray binary.

 
\end{abstract}

\section{Introduction} \label{intro}

The X-ray transient \grs\ was first discovered  by
Castro-Tirado et al. (1992) using the WATCH all-sky
monitor on board the Russian satellite GRANAT. 
Special attention to this source increased enormously
a few years later when Mirabel \& Rodr\'{\i}guez (1994)
reported that \grs\ was producing repeated ejections with apparent
superluminal motion.  These events could be first followed
at radio wavelengths thanks to the Very Large Array interferometer
of NRAO and represented a phenomenon never
observed before in the Galaxy. There are only two 
confirmed galactic superluminal sources known, the other one
being GRO J1655$-$40 (Hjellming \& Rupen 1995; Tingay et al. 1995).
Both of them have been classified as `microquasar' systems, i.e., a
selected group of radio emitting X-ray binaries whose physics and
external appearance is strongly reminiscent of quasars and active galactic nuclei.
We refer the reader to the review by Mirabel \& Rodr\'{\i}guez (1999) for
an updated account. 

\begin{figure*}[htb]
\mbox{}
\vspace{11.0cm}
\includegraphics{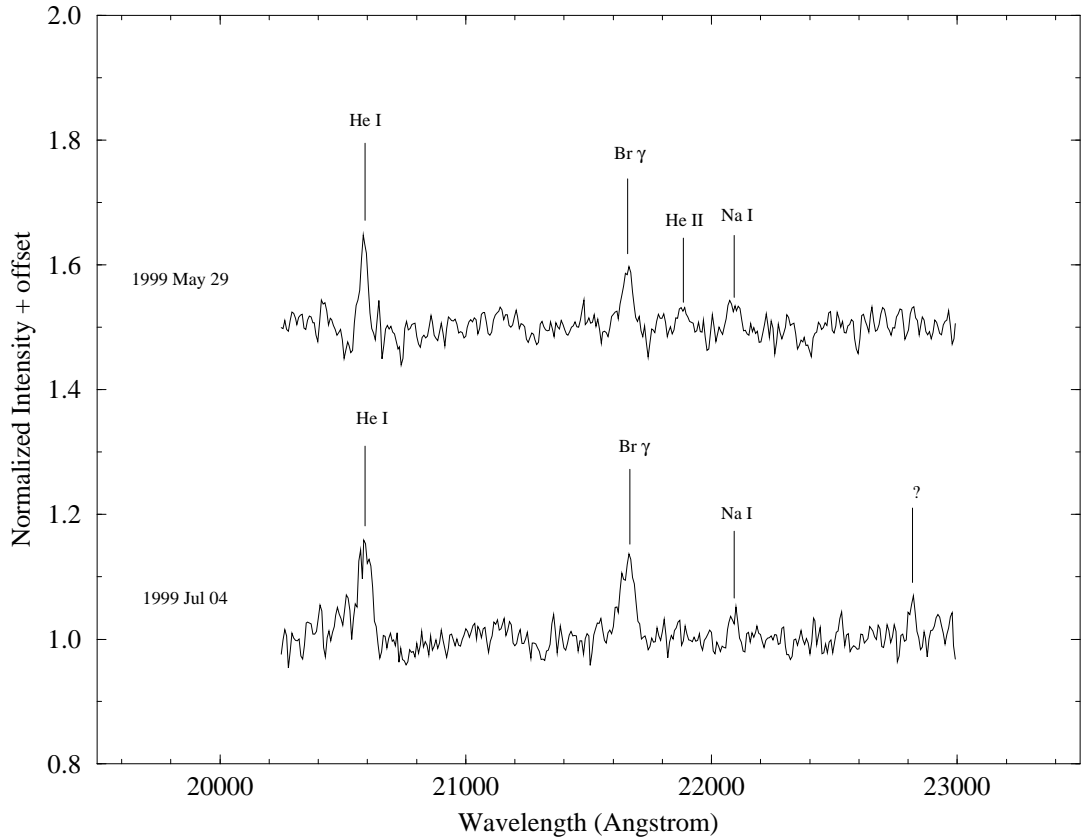}
\caption[]{VLT low resolution spectra of \grs\ obtained on 1999 May 29
(before a radio outburst) and
1999 July 04 (during a radio outburst) using the SK filter with ISAAC. The emission line identifications
are indicated.}
\label{isaac}
\end{figure*}

Since the discovery of superluminal episodes in \grs, this object
has been under strong surveillance by several observers at different
wavelengths. In spite of this work, the nature of the X-ray binary system
in \grs\ is still a matter of debate. The strong interstellar absorption
towards the source, that implies optical extinctions of $A_V \sim 27$ magnitudes
(Chaty et al. 1996), appears as the main cause of this uncertainty.
Therefore, it is not surprising that the best approaches to the problem
have come from imaging and spectroscopic studies at infrared wavelengths, where
the effects of extinction are much less severe ($A_K \sim 3.0$ mag). 
Infrared results, however, have
not yet reached an unanimous consensus. \grs\ has been proposed to be
a low-mass X-ray binary (LMXRB) by Castro-Tirado et al. (1996) and
an high-mass X-ray binary (HMXRB) by Chaty et al. (1996)
and Mirabel et al. (1997), who favour an Oe/Be spectral type. There is
also a discussion on the reality of weak emission lines of He II,
that have been marginally detected by Castro-Tirado et al. (1996)
and Eikenberry et al. (1998a). 

The existence, or not, of He II lines
is relevant because they are naturally expected in LMXRBs as a result of excitation from 
the hard photons of the accretion disk. The prototypical LMXRB Sco X-1 is a good example
of this statement (Bandyopadhyay et al. 1999).
In contrast, He II is seldom observed in HMXRBs. In massive systems,
accretion often proceeds directly from the stellar wind    
and photons from typical companion photospheres
would not be energetic enough to ionize helium atoms. However, the He II criterion
is not a reliable one to distinguish between high and low mass systems. Indeed, 
a few objects in the HMXRB group do exhibit He II lines in emission. The best
representative is certainly A0538$-$66 (Charles et al. 1983) in the Large Magellanic Cloud. 
This B2 III-IV system
has been observed to produce strong and variable He II 4648 \AA\ emission when in outburst.
An interesting property of some emission lines in A0538$-$66 (e.g. C IV and He II) 
is that they alternate between a P Cygni profile and a noticeable blue wing in emission when in quiescence and
outburst, respectively. We advance here that the same behavior may be present in \grs. 
Another case of massive He II emitter is the classical black hole candidate Cygnus X-1,
whose spectrum displays He II 4846 \AA\ emission believed to form in the stellar wind above the
substellar point of the O9.7 Iab primary (Gies \& Bolton 1986).

In this context, we present new infrared spectroscopic observations of \grs\ 
obtained with the first 8.2 m unit (Antu) of the Very Large Telescope (VLT)
of the European Southern Observatory (ESO).
The VLT spectra show, among others ones, possible
evidence for He II and a newly detected doublet of Na I, both in emission.
When taking all these features into account we find 
compelling evidence that a HMXRB scenario, with a surrounding envelope,
provides a very likely interpretation for \grs. 

\section{Observations and results} \label{obser}

\begin{figure*}[htb]
\mbox{}
\vspace{9.0cm}
\includegraphics{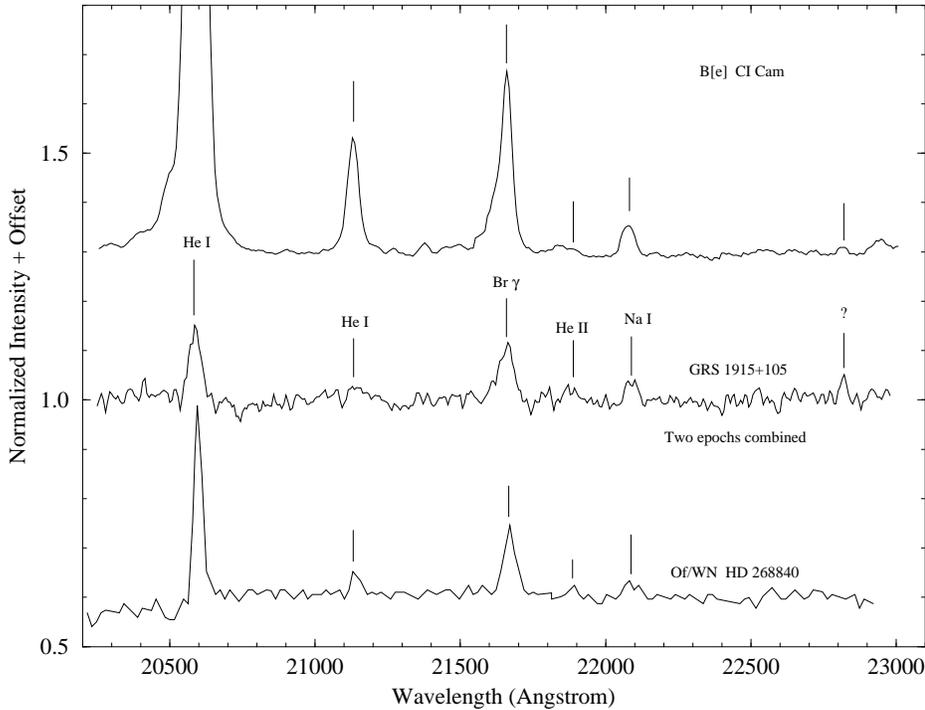}
\caption[]{Combined VLT spectrum of \grs\ (middle) obtained by averaging the data
of the two epochs to increase signal-to-noise ratio in order to better
detect any weak persistent features. The He II line in \grs\ appears barely
visible because it is not present in the second epoch, while 
the faint He I feature may have some contribution from N III. The two
comparison spectra are those of the B[e] system CI Cam (top)
and the Ofpe/WN9 star HD 268840 (bottom) adapted from Clark et al. (1999)
and Morris et al. (1996), respectively. The common features are identified on all plots
and labeled for \grs.}
\label{average}
\end{figure*}

We observed \grs, in service mode, on two different epochs using the ISAAC spectro-imager
at the VLT (see Cuby 1999 for a description). The first epoch was on 1999 May 29
(JD 2451327.84) under good weather conditions. The second observation
took place on 1999 July 04 (JD 2451363.77) under less favorable weather conditions.
In both dates, we carried out low resolution ($R \sim 450$) spectroscopy with the SK filter
and an integration time of 30 minutes.
The $1^{\prime\prime}$ slit was selected, with position angle of $90^{\circ}$
in order to avoid bright nearby stars. The frames were acquired
with the detector integration time (DIT) 
set at the recommended value of 60 s. Each observing epoch consisted of acquiring
a total of 30 of such frames. The usual technique of nodding the telescope was used, 
so that the target is placed at different slit positions in each frame.

The data were reduced using the IRAF package, including bias and dark current subtraction,
flat field division and frame combination. The wavelength calibration was
performed by means of the OH sky lines in the range 1.9-2.25 $\mu$m. 
The atmospheric absorption was corrected by observing the G4 V star HIP 99727, 
at an air mass as close as possible to that of \grs. This star was observed
by the telescope staff as a part of the ESO calibration plan. The final target spectrum
was obtained as the ratio of the absorbed spectrum to that of the solar type star.
The only intrinsic stellar absorption
present in the $K$-band, namely Br $\gamma$, was adequately removed by interpolation of the nearby continuum
prior to computing the ratio.


The main results of this paper are presented in Figs. \ref{isaac} and \ref{average}. 
The absorption correction worked
acceptably well for the first spectrum. The second one suffered from
less good atmospheric conditions and a slight 
mismatch (\ltsima 0.1) of air masses between the target and HIP 99727. 
The absorption correction close to the $\sim2.0$ $\mu$m region, with
strong telluric features, was consequently difficult but we still consider it acceptable.

The vertical axes in the spectra of Figs. \ref{isaac} and \ref{average} are expressed in terms
of relative flux plus an arbitrary offset. No absolute calibration could be
performed. This is because HIP 99727 does not have yet $K$-band magnitudes 
available in the literature. Nevertheless, a relative comparison of the
continuum level and line fluxes can actually be obtained. During the second epoch,
the continuum was $\sim6$\% brighter than during the first one. This implies 
a photometric variation of $\Delta K \simeq 0$\rmag 1, that may be consistent
with zero. In contrast, the emission line parameters of \grs\ do seem to have varied
significantly by a factor of $\sim2$. We derived them by fitting a Gaussian
profile and the corresponding results are also given in Table \ref{lines}. 
The likely error of the fits is estimated to be 
less than 1 pixel (\ltsima10 \AA) in wavelength
and $\sim10$\% in the relative flux, equivalent width (EW) and full width half
maximum (FWHM), respectively. The following subsections are devoted to comment
on the different emission lines listed and identified
in Table \ref{lines}.

\begin{table}
\begin{center}
\caption[]{\label{lines} Emission line parameters observed in \grs}
\begin{tabular}{cccccc}
\hline
Date         & Line                  & Observed $\lambda$  &  Relative  &   EW      &  FWHM     \\
1999         & Id. ($\mu$m)          &  ($\mu$m)           &  Flux      &   ( \AA ) &  ( \AA )  \\
\hline
May 29       &  He I       2.0587    &   2.0585            &    0.88    &  $-6.1$   &   36      \\
             &  Br$\gamma$ 2.1661    &   2.1658            &    1.00    &  $-5.5$   &   55      \\
             &  He II      2.1891    &   2.1886            &    0.49    &  $-2.7$   &   63      \\
             &  Na I  2.206-9        &   2.2087            &    0.57    &  $-2.9$   &   58      \\
             &                       &                     &            &           &           \\
July 04      &  He I       2.0587    &   2.0589            &    0.87    &  $-10.8$  &   67      \\
             &  Br$\gamma$ 2.1661    &   2.1658            &    1.00    &  $-10.0$  &   73      \\
             &  He II      2.1891    &   $-$               &    $-$     &  $-$      &   $-$     \\
             &  Na I  2.206-9        &   2.2087            &    0.24    &  $-2.2$   &   42      \\
             &  ?                    &   2.2817            &    0.29    &  $-2.4$   &   32      \\
\hline
\end{tabular}
\end{center}
\end{table}

\subsection{Apparent P Cygni and asymmetric emission profiles in the He I and Br $\gamma$ lines}

The infrared spectra for the two epochs in Fig. \ref{isaac}
confirm the detection of the He I (2.0587 $\mu$m) and Br $\gamma$ (2.1661 $\mu$m) emission lines, both without
significant Doppler shifts. 
These lines have been reported before on different epochs by
Castro-Tirado et al. (1996), Mirabel et al. (1997) and Eikenberry et al. (1998a).
The line ratios He I/Br$\gamma$ seem to be often 
in the range 0.96-1.16 as derived from these previous detections.
In both our two VLT spectra we find that He I/Br $\gamma$ is practically 0.9, i.e., a value 
not far from this range.

On the first epoch, He I appears to display a P Cygni profile corresponding to an expansion velocity
of $\sim2000$ km s$^{-1}$. The same type of profile may also be present in Br $\gamma$, although it is not so obvious.
In contrast, during the second epoch we find that the He I and Br $\gamma$ lines have changed
their appearance to a noticeable asymmetric profile. The asymmetry is particularly evident in He I. 
A blue wing component in emission seems to have emerged in {\it both lines} also 
extending up to $\sim2000$ km s$^{-1}$. 
Hereafter, we will proceed under the assumption
that this is a real effect because the two lines appear to show it.
Furthermore, the skewness of the He I profile for 
1999 May 29 is $-0.35\pm0.27$, consistent with
no significant asymmetry, while that of
1999 July 04 gives a skewness of $-0.73\pm0.14$,
indicating the presence of a blue asymmetry significant
at the 5-$\sigma$ level.
Nevertheless, it will be interesting to obtain future confirmation of it that safely rules out
any possible small error when correcting for atmospheric transmission.  

\subsection{Confirmation of He II emission}

In our first epoch spectrum, we detect a weak feature in emission close to
2.1891 $\mu$m. This is the expected wavelength of the He II 10-7 transition
marginally detected by Castro-Tirado et al. (1996) and Eikenberry et al. (1998a).
The plausible He II/Br $\gamma$ ratios derived from their spectra are about 0.53 and 0.23, 
respectively. We find this ratio in our case to be about 0.49, hence well within this range.
The He II emission is, once more, detected at a marginal level. However, the fact that this is
the third independent marginal detection gives us confidence about its reality.


\subsection{Other emission lines}

The two VLT spectra seem also to contain a feature in emission centered at 2.2087 $\mu$m. 
This could be consistent with the Na I doublet at the wavelengths of
2.206-9 $\mu$m. The significance of this detection improves noticeably if one
ave\-rages the two individual spectra, as we show in Fig. \ref{average}.
This combined plot also contains some possible evidence for the complex of He I
lines located at 2.112-3 $\mu$m. At the resolution and signal-to-noise ratio of our spectra,
it is difficult to discriminate if this marginal feature
could also be due to N III at the nearby wavelength of 2.116 $\mu$m.
Finally, an unidentified emission feature at 2.2817 $\mu$m is apparent
in the 1999 July 04 spectrum.  

\section{Multi-wavelength behaviour of \grs\ during the VLT runs}

\begin{figure}[htb]
\mbox{}
\vspace{9.0cm}
\includegraphics{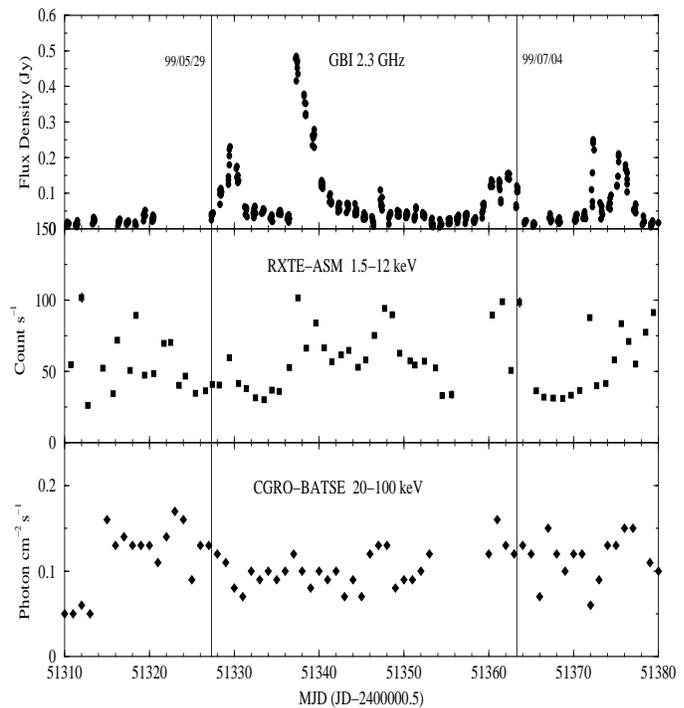}
\caption[]{{\bf Top.}
The radio behavior of \grs\ during our
observations as recorded by the daily GBI monitoring.
The vertical lines indicate the precise dates when the VLT-ISAAC spectra
were obtained. {\bf Middle.} The same kind of plot
but showing the soft X-ray light curve of \grs\
from the RXTE-ASM data. {\bf Bottom.} Another similar plot
with the hard X-ray light curve of \grs\ from the CGRO-BATSE Earth occultation monitoring.
}
\label{moni}
\end{figure}

It will be useful for discussion purposes to provide an overview 
of the X-ray and radio behaviour of \grs\ at the times of our VLT observations.
This information is displayed in the panels of Fig. \ref{moni}. From top to bottom, 
we plot here the 2.3 GHz, 1.5-12 keV and 20-100 keV light curves of \grs.
These multi-wavelength data have been obtained thanks to the public monitorings
available from the Green Bank Interferometer (GBI), 
the All Sky Monitor (ASM) on board the Rossi X-Ray Timing Explorer (RXTE) and
the Burst and Transient Source Experiment (BATSE) on board the Compton Gamma-Ray Observatory (CGRO),
respectively.

We see in Fig. \ref{moni} that the first epoch of
observation was just prior to the onset of an
ejection event, with a $\sim0.25$ Jy amplitude in
the radio, and a minor soft X-ray maximum.
The second epoch was obtained soon after another strong ejection  
event had taken place. Its amplitude was similar to that of the May event
and it also had a noticeable soft X-ray counterpart. The second run of VLT observations 
coincided with the early decay of the flare.
It has been proposed that the ejections in \grs\ involve the
destruction and ejection of the inner accretion disk and corona, as evidenced by previous multi-wavelength
observations (Mirabel et al. 1998; Fender \& Pooley 1998; Eikenberry et al. 1998b).

\section{Discussion}

\subsection{Interpretation of P Cygni and asymmetric wing emission profiles}

The apparent P Cygni profile detected on 1999 May 29 (see Fig. \ref{isaac}) corresponds to an epoch when \grs\
was not yet in outburst. This profile is naturally understood in terms of
an expanding shell or extended atmosphere around the companion, whose approaching (and therefore blueshifted)   
regions are seen in absorption against the stellar disk. 
It is appropriate to remind here that evidences of surrounding material around \grs\
have been reported in the past (Mirabel et al. 1996).

Several weeks later, the $\sim0.25$ Jy outburst that peaked $\sim2$ d before our second 
spectrum seems to have altered considerably the quiescent picture. The shell around   
\grs\ is likely to have experienced a considerable injection of energy
from the outburst, with part of its material being blown out. It is this
blown out material what we believe is producing the strong blue emission
wing in the July 04 spectrum. The evolution from quiescent P Cygni profiles to blue emission wings
is, in fact, not a new phenomenon in the context of X-ray binaries. The same behavior has
been very well observed in other outbursts of different sources. As mentioned in the
introduction, the B2 III-IV system A0538$-$66 is one of such cases  
(Charles et al. 1983). A more recent example can be found in the 1998 outburst of
CI Cam, which has been classified as a B[e] 
HMXRB by Clark et al. (1999). Their $K$-band spectra of CI Cam in outburst 
clearly shows the blue wing effect, specially in the He I and Br $\gamma$ lines (see Fig. \ref{average}). 

Further evidence of blue wing emission, with superimposed P Cygni profiles, is also present
in the $K$-band spectrum of \grs\ by Mirabel et al. (1997) of 
1995 September 4. The blue asymmetry in the He I
line in Fig. 1 of this paper is even more pronounced.
The GBI monitoring was not available at that time, but the 
contemporaneous BATSE light curve displays a sharp drop $\sim3$ d before that date,
and we know that the sudden drops in the BATSE flux are associated to major ejection events.

\subsection{The HMXRB versus LMXRB scenarios}

One of the important results provided by the VLT spectra is an independent
confirmation of the weak He II emission line, at least on the first
epoch spectrum. We do not have a reliable detection of He II on the second epoch, although this
could be due either to line variability or to the lower quality of the night.

As mentioned before, the debate on the origin of the
He II line has important consequences on the nature of the \grs\ companion.
Since no ordinary star is able to generate it,
Castro-Tirado et al. (1997) and Eikenberry et al. (1998a) 
attribute its detection to the accretion disk around the compact object 
rather than the atmosphere of a Be star or other massive companion.
However, when doing so another interesting possibility is excluded. He II emission can  
actually be excited in the atmosphere of the hottest stars in the spectral sequence.
For instance, the spectra of some O-type stars are well known to exhibit He II both in absorption
and emission. There is actually a sequence between pure absorption O stars and stars with both
He II and N III in emission, that are known as Of stars (Kaler 1989). 
Therefore, the detection of He II emission does not necessarily rule out that we
are dealing with a HMXRB in \grs\ provided that the primary is of very early type.
 
The challenging HMXRB possibility, with an early type companion, fits nicely with the
expanding envelope evidences based on P Cygni and asymmetric wing profiles.  
The idea also appears further supported when comparing our VLT spectra with those of massive stars 
(see Fig. \ref{average}).
Excellent samples of spectra of luminous stars belonging to the classes of Of, Wolf-Rayet (WR), Be, 
B[e] and Luminous Blue Variables (LBV) can be found, e.g., in the works by Morris et al. (1996), 
Hanson et al. (1996) and Clark \& Steele (1999). The discrimination between these different classes of stars
on the basis of 2 $\mu$m spectroscopy is not always straightforward, because of significant
overlap in their spectral morphology.
Individual stars have also been
observed to change their spectral morphology with time and move from one
class to another on timescales of a few years. 
Our situation is further worsened here due to the low signal-to-noise ratio of the spectra.
However, even with these limitations, it should be still possible to recognize the HMXRB nature
of \grs\ and to provide a tentative spectral classification
from comparison with spectral atlases of hot, luminous stars in the infrared. 

With these cautions into account let us consider the contents of 
Table 8 in the Morris et al. (1996) paper, where the
different spectral characteristics of massive stars in the $K$-band are listed.
We find here that the observed properties of 
Br $\gamma$ emission with $EW\sim10$ \AA, as well as emission lines of
He I at 2.112-3 $\mu$, N III at 2.116 $\mu$m,
He II at 2.1891 $\mu$m and Na I at 2.206-9 $\mu$m are simultaneously present
only for stars of the Of and WR nitrogen (WN) classes. Interestingly, this seems to
be in good agreement with the most likely spectral features observed or suspected in our VLT spectra.

The presence of Na I implies that part of the stellar wind of the
proposed Of/WN companion has to be shielded
from direct stellar radiation because of the low ionization potential of sodium, perhaps in a cooler
dusty region of the star atmosphere. Na I is also a typical feature in the 
related HMXRBs of the Be type (Everall et al. 1993).
Concerning the strong He I line at 2.0587 $\mu$m, it is often excluded as
a classification criterion because of several factors that make it unreliable 
(difficult telluric correction, optical depth variability, bad indicator to distinguish
between OB and WR subtypes, etc.). He I in emission is believed
to be related to luminosity class I, but this is a rule with several known exceptions. 

Another issue that strengthens the relationship between \grs\ and massive luminous stars is
again provided by the comparison of our spectra with those of CI Cam.
As \grs, CI Cam has been observed to undergo violent ejection events that are
able to produce bipolar collimated jets with relativistic velocities (Hjellming \& Mioduszewski 1998). 
The evidence so far available strongly suggests that CI Cam may be another member of 
the microquasar group. It is then appropriate to remind here that 
the B[e] stars in the infrared are sometimes hard to distinguish from the Of/WN status 
and viceversa, e.g., as discussed by Morris et al. (1996) in the cases of
WR 85a and WR 122.
With these thoughts in mind, the infrared spectrum of CI Cam by Clark et al. (1999) 
compares remarkably well with our \grs\ spectrum as evidenced in Fig. \ref{average}. 
All of them are coincident again at displaying emission from  
Br$\gamma$, He II 2.1891 $\mu$m, the Na I doublet and the He I 2.0587 $\mu$m line.
From this comparison, it appears very plausible to us that both \grs\ and CI Cam are HMXRBs 
in closely related evolutionary stages.

\subsection{The He II line and the accretion disk}

The fact that a very massive early-type companion can produce the He II emission
does not exclude the possibility that the accretion disk is also contributing to this feature.
It is conceivable as well that the recurrent creation and disruption of the inner accretion
disk during the ejection events is responsible for most of the observed variability of this line.
The fact that we better detect enhanced He II emission
on the first epoch could be understood considering  
that \grs\ was then prior to a strong ejection event (see Fig. \ref{moni}). Therefore, the accretion
disk at that time should be well extended up to the latest stable orbit. 
In contrast, the second epoch almost coincided
with the peak of another ejection episode (see again Fig. \ref{moni}). 
The inner accretion disk from which the ejecta is formed
may have not been yet refilled, hence the lower He II emission.

\subsection{HMXRBs among galactic radio jet sources}

To complete our discussion, it is interesting to point out that there
seems to be a majority of HMXRBs or intermediate mass systems among the list of
confirmed radio jet sources in the Galaxy. 
This selected group includes with great confidence 
SS 433, Cyg X-3, Cir X-1 and GRO J1655$-$40. 
The first three of these objects are widely believed to be 
bona fide HMXRBs. The last one has been reported be a system of intermediate
mass, with the optical companion being in the range 1.7-3.3 $M_{\odot}$ (Shahbaz et al. 1999).
Add the recently discovered massive jet source CI Cam
and the idea of \grs\ being also a HMXRB does not come with great surprise. 

However, given that \grs\ is a remarkable
binary system with a very luminous accretion disk
and relativistic ejections, other origins for the observed
lines different from a massive companion cannot be strictly ruled out at present. 
In particular, the fast variability in the infrared emission
lines observed by Eikenberry et al. (1998a) is difficult to explain 
with a normal stellar atmosphere, unless it is being directly affected by
the X-ray flares and jets. Indeed,
\grs\ is a complex object where interactions between the  
massive outflows from the companion star and relativistic 
ejections from the compact object are likely to occur.
In any case, additional studies on the nature of the companion of the collapsed object should
be pursued.

\section{Conclusions}

\begin{enumerate}

\item We have presented the first VLT infrared spectra of the superluminal source \grs.
The strong emission line of He I at 2.0587 $\mu$m provides suggestive evidence of
a P Cygni profile evolving into a blue wing in emission when the system goes from
quiescence to outburst. A similar trend is also likely to occur for Br $\gamma$. 
This behavior is consistent with what is observed during the outbursts of other 
HMXRBs and points towards the existence of an expanding shell or atmosphere around
the \grs\ companion. We speculate that the appearance of such shell 
is changing depending on the state of activity of the high energy source. 

\item The \grs\ companion is very likely to be a massive star.
Not only the presence of an extended envelope is a clue of its HMXRB
nature, but also the comparison of the \grs\ spectrum with infrared spectral atlas
supports this interpretation. The newly discovered line features, specially Na I, as well as those
previously detected show a strong resemblance with the spectra of such massive stars. 
We are also able to propose a tentative spectral classification of the
primary as an Of/WN star. 

\item In this context, the long discussed He II line should be excited by the photons
of the massive and luminous companion star, although we cannot rule out completely
that the accretion disk is also playing a role. 
In any case additional sensitive observations, in different states
of activity of the source, would be very valuable to settle definitively
this issue and to provide a more accurate spectral type.

\end{enumerate}

\begin{acknowledgements}
We are very grateful to P.A. Charles (Univ. of Oxford) for helpful discussions on the
interpretation of our infrared spectra.  We also thank
F. Comer\'on (ESO) and R. Zamanov (National Rozhen Observatory) for 
advice in the reduction of the VLT data, and 
J.S. Clark (University of Sussex) for kindly providing the CI Cam comparison spectrum.
J.M. acknowledges partial support by DGICYT (PB97-0903)
and by Junta de Andaluc\'{\i}a (Spain).
S.C. acknowledges financial support from the Leverhulme foundation.
This paper is partially based on quick-look results provided by the
ASM/RXTE team and data obtained through the HEASARC Online Service of
NASA/GSFC.  The GBI is a facility
of the USA National Science Foundation operated by NRAO in support of the
NASA High Energy Astrophysics programs. 

\end{acknowledgements}


\begin{thebibliography}{}

\bibitem[1992]{bandy}
Bandyopadhyay R.M., Shahbaz T., Charles P.A., et al., 1999, MNRAS 306, 417

\bibitem[1992]{castro92}
Castro-Tirado A.J., Brandt S., Lund, N., 1992, IAU Circ. 5590

\bibitem[1992]{castro96}
Castro-Tirado A.J., Geballe T.R., Lund N., 1996, ApJ 461, L99 

\bibitem[1983]{charles83}
Charles P.A., Booth L., Densham R.H., et al., 1983, MNRAS 202, 657

\bibitem[1996]{chaty}
Chaty S., Mirabel I.F., Duc. P.-A., et al., 1996, A\&A 310, 825

\bibitem[1999]{clark1}
Clark J.S., Steele I.A., 1999, A\&A (in press)

\bibitem[1999]{clark2}
Clark J.S., Steele I.A., Fender R.P., Coe M.J., 1999, A\&A 348, 888

\bibitem[1999]{cuby}
Cuby J.G., 1999, {\it ISAAC User Manual}, ESO

\bibitem[1998]{eikena}
Eikenberry S.S., Matthews K., Murphy T.W., et al., 1998a, ApJ 506, L31

\bibitem[1998]{eikenb}
Eikenberry S.S., Matthew K., Morgan E.H., et al., 1998b, ApJ, 494, L61

\bibitem[1993]{everall}
Everall C., Coe M.J., Norton A.J. et al., 1993, MNRAS 262, 57



\bibitem[1998]{fender98}
Fender R.P., Pooley G.G., 1998, MNRAS 300, 573

\bibitem[1986]{gies}
Gies D.R., Bolton C.T., 1986, ApJ 304, 371

\bibitem[1996]{hanson}
Hanson M.M., Conti P.S., Rieke M.J., 1996, ApJSS 107, 281

\bibitem[1995]{hj1}
Hjellming R.M., Rupen M.P., 1995, Nat 375, 464

\bibitem[1998]{hj2}
Hjellming R.M., Mioduszewski A.M.,  1998, IAU Circ. 6872


\bibitem[1989]{kaler}
Kaler J.B., 1989, {Stars and their spectra}, Cambridge University Press

\bibitem[1994]{mr94}
Mirabel I.F., Rodr\'{\i}guez L.F., 1994, Nat 371, 46

\bibitem[1996]{mr96}
Mirabel I.F., Rodr\'{\i}guez L.F., Chaty S., et al., 1996, ApJ 472, L111

\bibitem[1997]{mirabel}
Mirabel I.F., Bandyopadhyay R., Charles, et al.,
1997, ApJ 477, L45


\bibitem[1998]{mirabel98}
Mirabel I.F., Dhawan V., Chaty S., et al., 1998, A\&A 330, L9

\bibitem[1999]{mr1999}
Mirabel I.F., Rodr\'{\i}guez L.F., 1999, ARAA 37, 409 

\bibitem[1996]{morris1996}
Morris P.W., Eenens P.R.J., Hanson M.M., et al., 1996, ApJ 470, 597

\bibitem[199]{shabaz}
Shahbaz T., van der Hooft, Casares J., et al., 1999, MNRAS 306, 89

\bibitem[1995]{tingay}
Tingay S.J., et al., 1995, Nat 374, 141


\end{thebibliography}
\end{document}